# Domain-topic models with chained dimensions: charting an emergent domain of a major oncology conference

Alexandre Hannud Abdo[1,5], Jean-Philippe Cointet[2], Pascale Bourret[3], Alberto Cambrosio[4]

[1] LISIS, Univ Gustave Eiffel, ESIEE Paris, CNRS, INRA, F-77454 Marne-la-Vallée, France
[2] Sciences Po, médialab, Paris, France
[3] Aix Marseille Univ, INSERM, IRD, SESSTIM, Marseille, France
[4] Department of Social Studies of Medicine, McGill University, Canada
[5] Garoa Hacker Clube, São Paulo, Brazil

# Abstract

This paper presents a contribution to the study of bibliographic corpora in the context of science mapping. Starting from a graph representation of documents and their textual dimension, we observe that stochastic block models (SBMs) can provide a simultaneous clustering of documents and words that we call a domain-topic model. Previous work by (Gerlach et al., 2018) investigated the resulting topics, or word clusters, while ours focuses on the study of the document clusters, which we call domains. To enable the synthetic description and interactive navigation of domains, we introduce measures and interfaces relating both types of clusters, which reflect the structure of the graph and the model. We then present a procedure that, starting from the document clusters, extends the block model to also cluster arbitrary metadata attributes of the documents. We call this procedure a domain-chained model, and our previous measures and interfaces can be directly transposed to read the metadata clusters. We provide an example application to a corpus that is relevant to current STS research, and an interesting case for our approach: the 1995-2017 collection of abstracts presented at ASCO, the main annual oncology research conference. Through a sequence of domain-topic and domain-chained models, we identify and describe a particular group of domains in ASCO that have notably grown through the last decades, and which we relate to the establishment of "oncopolicy" as a major concern in oncology.

# 1. Introduction

Within the tradition of co-word analysis (Callon et al., 1983), understood as the first attempt at using the content of documents to capture the dynamics of scientific activity, a variety of methods have been adopted to reveal meaningful relationships between words, documents, and other dimensions in a corpus, among which are semantic maps and topic models (Chavalarias & Cointet, 2013; Hecking & Leydesdorff, 2019). Recent work (Gerlach et al., 2018) has shown that a family of network models, called Bayesian stochastic block models (Peixoto, 2018), offers an interesting topic modeling alternative to established LDA models (Blei et al., 2003). In the present paper we explore the fact that these network models can be employed to simultaneously infer document clusters, and we introduce procedures and tools to systematically interpret these clusters in combination with the topic model. We also introduce a method to extend this approach to other dimensions, thus covering a range of applications such as period detection and author topic models (Rosen-Zvi et al., 2004).

The approach presented here represents an attempt to avoid the compromises of lacking a comprehensive statistical formulation, as is the case with semantic maps, or treating documents as elements of a flat landscape, as in topic models. If topic models group words according to patterns in their occurrence in documents, forming topics, our approach also accounts for patterns in documents' usage of words, gathering documents into groups we call the domains of a domain-topic model. Other procedures that combine document clustering and topic modeling have been proposed, all based on stacking variants of LDA with models that produce document clusters, ranging from the naive application of k-means on top of a topic model to more thoughtful, composed models such as MGCTM [(Xie & Xing, 2013)](Xie & Xing, 2013). Our goal in this paper is not to quantitatively compare our results to those procedures, but to highlight and explore the qualitative possibilities and tools that follow from the simplicity and flexibility of the present approach.

Among its advantages, we show that the resulting lexically structured document landscape can be straightforwardly used as a lens to cluster and read other dimensions that converge in a document, such as metadata, through what we will call domain-chained models. And, like co-word maps, which provide easy to navigate lexical cartographies, we are able to translate our models into interactive maps that visually tie clusters of documents with their lexical and metadata dimensions, allowing for the navigation and description of the corpus at multiple scales and aspects.

As a demonstrator for our approach, we pick a concrete research object in the sociology of science. We investigate the abstracts of the annual meetings of the American Society for Clinical Oncology (ASCO) from 1995 to 2017, and show how we can lay out 23 years of the world's largest oncology conference in terms of research domains evolving through a sequence of periods where different domains rise and fall. Then, we proceed to study in detail one of the notable shifts, showing it represents the rise of "oncopolicy" as several associated domains, a central issue nowadays as ASCO officially transitions from being a "mostly research" to being a "research and policy debate" organisation (ASCO, 2020).

Our choices aim to advance quantitative analyses that augment and empower qualitative exploration and reasoning, and belong in a trend to represent academic artifacts as hypergraphs, as recently discussed in (Cambrosio et al., 2020). They therefore call for a choice of statistical model that embodies an easily interpretable representation, and that is sufficiently generic and robust to treat different types of relationships, those of documents, words and metadata, leading us to adopt the specific SBM framework to be discussed. During the preparation of this paper, Gerlach independently published (Gerlach et al., 2018) closely related work that covers some techniques discussed here, and demonstrated the validity of this approach as a topic model. Our work thus extends Gerlach's by exploring the dimension of document clusters, their interpretation, and by extending the model towards arbitrary dimensions expressed in the metadata.

# 2. Methods

## 2.1 Domain-topic models

We consider that research domains can be defined as sets of scientific texts addressing the same questions, using shared methods, and focusing on the same or related entities, all of which get reflected in the terms expressed in those texts. Our approach thus first relies on the content of documents to produce a simultaneous clustering of documents and terms, where documents are organized in domains, whose documents from similar combinations of topics, while terms are organized in topics, whose terms have a similar presence among domains.

To analyze documents we adopt a classic bag-of-words hypothesis (Harris, 1954) and model each document as an unordered set of terms. As exemplified in figure 1, document 1 is composed by the terms: *"the"*, *"patient"* and *"surgery"*; document 2 by the terms: *"the"*, *"patient"*, *"average"*, and *"radiation"*; document 3 by the terms *"the"*, *"average"*, *"therapy"*, and *"breast_cancer"*.

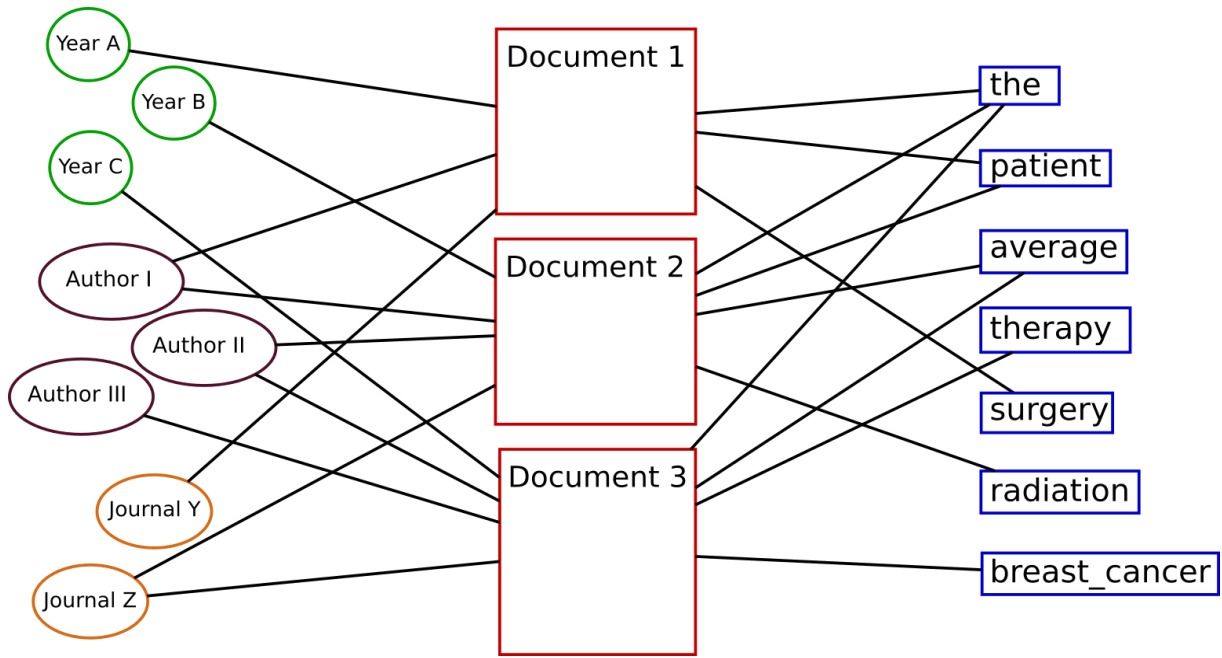

Figure 1: Incidence graph representation of relationships found in a research corpus. Each document appears as a node, with edges towards its textual content nodes (terms) and metadata nodes (i.e. authors, years, journals). Our goal is to study the corpus through the lens of its lexical dimension (terms), by clustering this graph restricted to the documents and term nodes (figure 2). Then, to use the lexically clustered documents to induce clusterings on the metadata dimensions they connect to (figure 3).

In a graph representation, each document is thus connected to its terms, and also to its metadata dimensions, which in Figure 1 are: year, authors and journal. This graph can be understood as the incidence graph of a hypergraph, whose nodes are the terms and metadata, and where documents are hyperedges associating multiple nodes. It is worth noting that an incidence graph is a bipartite graph: documents only connect to other types of nodes, and those only connect to documents.

In order to produce a domain-topic model, we restrict this graph to its document and lexical dimensions, shown in figure 2. We can then simultaneously cluster documents and terms to produce a categorization of both partitions: documents are organized in domains, while terms are organized in topics, as depicted by the brackets in the figure. In fact, figure 2 portrays a nested organization, with two levels of domains and topics, where clusters on the first level get clustered themselves, following the statistical model we will opt for.

Usually, content-based clustering of documents groups them into domains that share a similar term usage, whereas topic modeling groups terms into topics that share a similar presence in documents. A domain-topic model, which is proposed here, does both things at once in a synergistic way, resulting in a dual clustering, of domains on one side, and topics on the other, according to their connectivity patterns in the document-term graph.

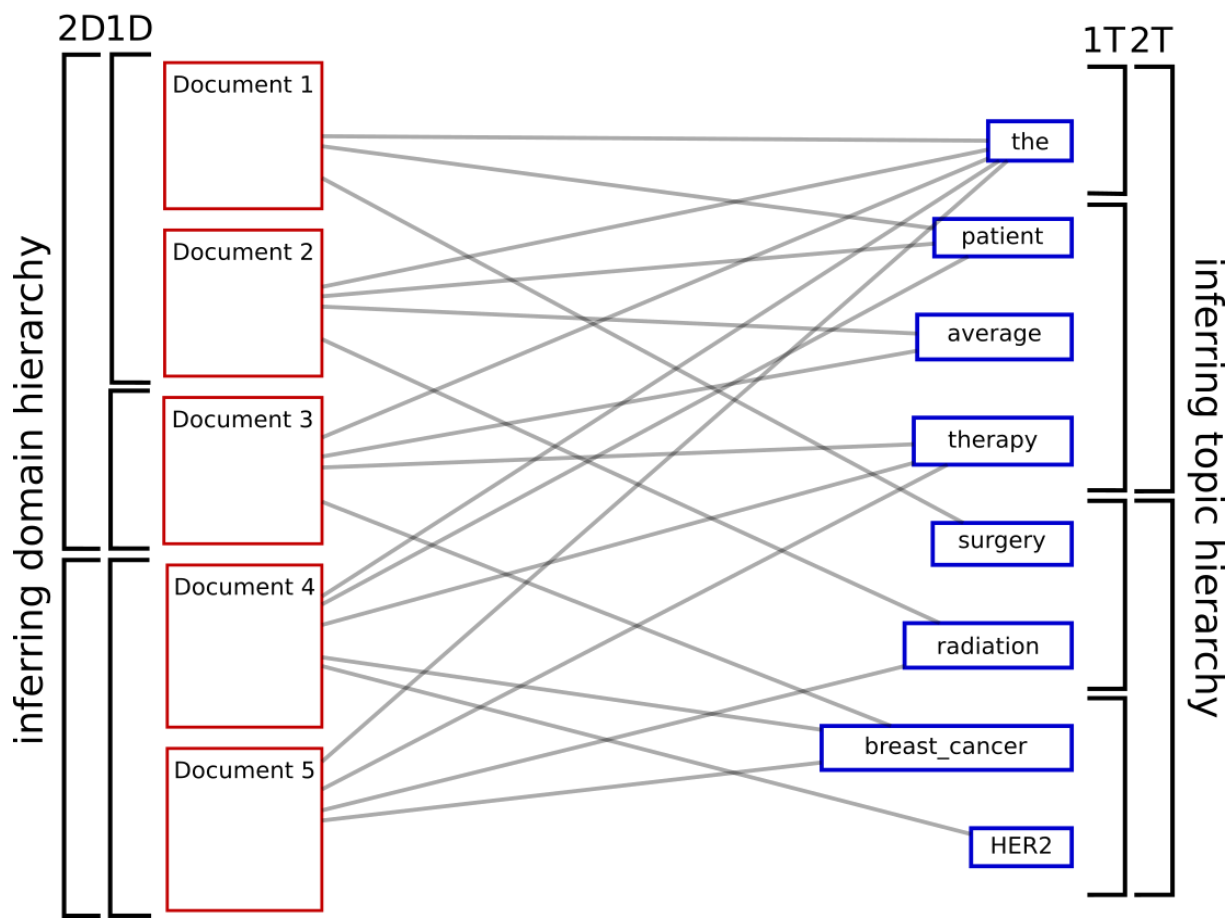

Figure 2: Domain-topic model of the bipartite graph of documents linked to their terms. The resulting dual structure features on the right side a hierarchy of topics (groups of terms) and on the left side a hierarchy of domains (groups of documents). Labels 1D = "level-1 domains", 2D = "level-2 domains", and similarly for 1T and 2T as topics.

One could, in principle, use any graph clustering method to deal with this task. However, the empirical difference in connectivity patterns on the two sides of this bipartite network raises the following issue. On the one hand, terms may vary from being present in only a few documents to spanning all of them, while, on the other hand, the number of terms per document in a corpus will fluctuate around some average. Put differently, in a typical document-term graph, the degree distribution of terms is wide and fat-tailed, while the degree distribution of documents is narrow. This suggests that we will either need to couple two separate models, or employ a model that can capture quite general patterns. That we will later include metadata, of an arbitrary nature, in our approach strongly points at the latter option.

Given that, in this paper we adopt as our graph clustering model the Bayesian Stochastic Block Model (Peixoto, 2014b), which as required clusters nodes according to general connectivity patterns, and provides a parsimonious method for model fitting and selection. Specifically, we adopt its nested, degree-corrected variant. In Stochastic Block Model (SBM) terminology, node clusters are called *blocks*. The basic SBM is a generative model where nodes are organized into blocks and connected according to edge probabilities between those blocks. Degree-correction improves on that by accounting for heterogeneous degree frequencies within blocks. And the Nested quality means it expresses patterns at different scales, by clustering blocks themselves into higher-level blocks, and so forth, forming a nested hierarchy. This stochastic model and related variants have been successfully applied to the analysis of both static and time-varying graphs, and have been shown to robustly reveal unexpected connectivity patterns while avoiding overfitting (Peixoto, 2015b).

In this SBM framework, fitting the model to network data is done by partitioning (grouping) the graph's nodes into blocks. For the nested model, this includes further partitioning the blocks themselves into higher-level blocks for each nested level. Given these partitions, blocks' degree sequences and block to block connection probabilities are simply traced from the edges in the graph. A model that best fits the graph is then searched following a minimal description length (MDL) approach, by seeking a partition that minimizes the combined informational costs of describing the graph given the model, and of describing the parameters we're selecting (Peixoto, 2014b). Notably, both the number of levels in the hierarchy and the number of blocks at each level get inferred from the data through the minimal description length principle.

This class of models provides our domain-topic model with a set of desirable properties:

- They detect *general connectivity patterns*; resolving the issue of treating different types of nodes: documents, terms and metadata.
- They are *non-parametric*; in particular, both the number of blocks and levels in the hierarchy are directly inferred from the data.
- They do *not overfit:* by accounting for the information cost of model parameters when maximizing model probability, they only infer statistically significant structures.
- They provide a nested, multi-level abstraction of terms and documents, allowing the investigation of topics and domains at different scales.

Another major concern in the study of corpora is the ease and reproducibility of result interpretation. This concern guides the following two additional choices regarding our modeling. First, stochastic models can be employed either by searching for a single model that best fits the data or by averaging the values of interest over a distribution of models yielded according to their fitness. In this paper we choose to work with the single best-known fit for our data. Second, the model adopted allows for overlapping as well as non-overlapping blocks. In this paper we have worked with non-overlapping blocks. Contrary to what one might expect, it has been shown that non-overlapping models are often a better fit than overlapping ones (Peixoto, 2015a). This remains an area for future work, including the development of models that overlap topics but not domains, to better take into account the polysemy of terms. Together, these two choices greatly simplify interpretability insofar as each document is attached to a single domain, both probabilistically (single best fit) and concretely (non-overlapping).

In conclusion, we obtain our domain-topic models by fitting the above choice of SBM to a document-term graph, as pictured in figure 2. We perform our optimization procedures using the SBM implementation found in the graph-tool library (Peixoto, 2014a), which employs an efficient Markov Chain Monte Carlo (MCMC) approach. Our usage of the library and the specific procedures adopted for this paper are detailed in the supporting information (in SI-DATA).

## 2.2 Chained dimensions

Research domains also carry social and contextual dimensions, given that they are grounded in the activities of specific teams and institutions at a given time, whose results are made public at professional conferences and in scholarly journals. Generally regarded as metadata, these additional dimensions such as authorship, institutional affiliations, funding sources, or year of publication can be studied as a network in a number of different ways, but here we are interested in looking at them through the lens of the contents of documents. Meaning we want to cluster metadata elements that connect similarly across the domains of a domain-topic model.

To achieve that, after having inferred a domain-topic model for a corpus, we can use its domains to form an *inference chain* towards the documents' metadata. This is achieved by restricting the original graph (figure 1) to some metadata dimension, with documents connecting to their metadata nodes, as can be seen in figure 3. We then fit the SBM to this new graph, but transposing and keeping immutable the nested blocks of documents previously inferred by the domain-topic model. This way, only the metadata nodes get partitioned, as represented by the brackets on the left side of figure 3, and the best fit partition will reflect their connectivity patterns to documents and the domain hierarchy. We call this procedure a domain-chained model.

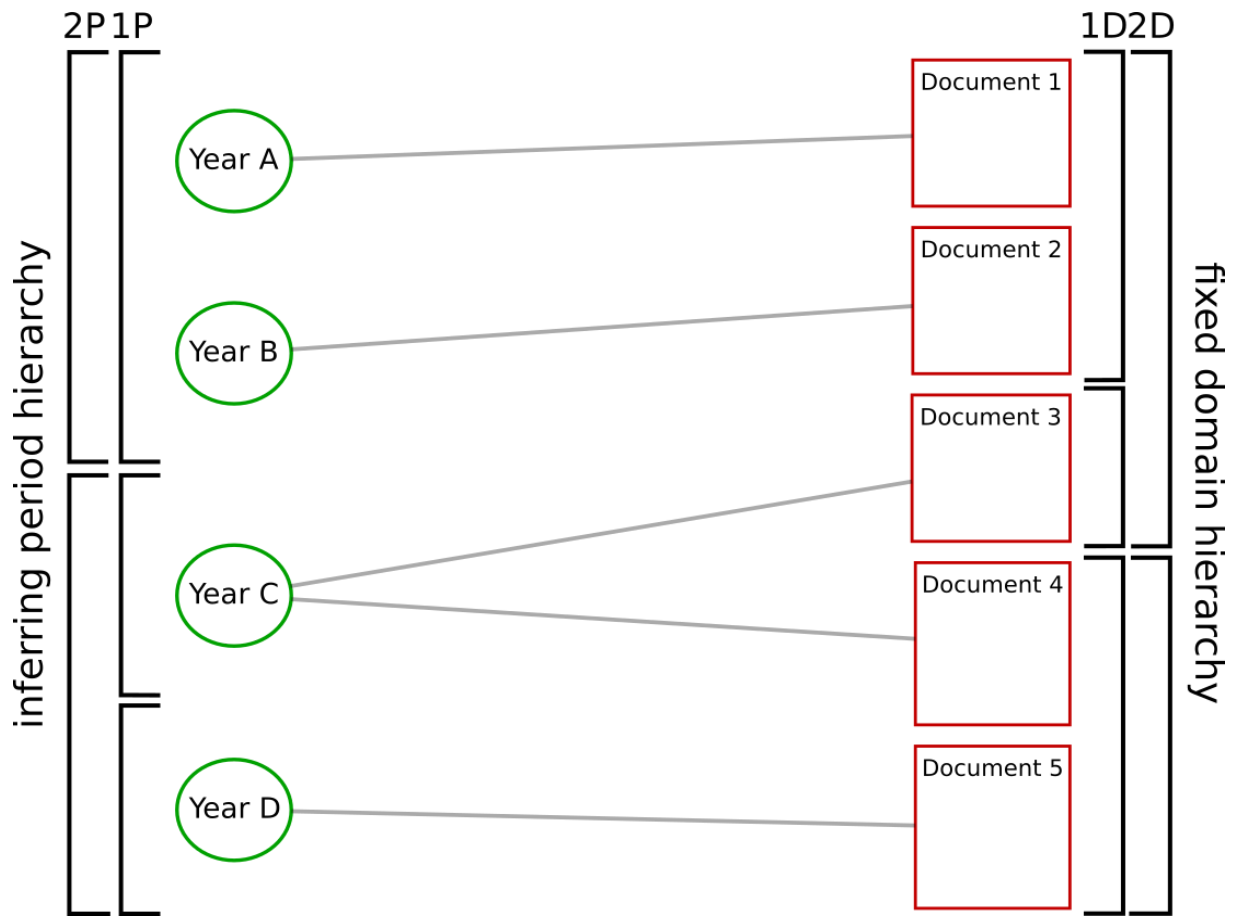

Figure 3: Domain-chained model of documents linked to their year of publication. By keeping the inferred document partition fixed, the model can be extended to other variables that get assigned to documents. In this exemple, publication years get partitioned into nested blocks. The best fit partition will reflect the connectivity patterns between the chained dimension and the lexically structured domains (1P = level-1 periods, 2P = level-2 periods).

## 2.3 Domain tables and interactive maps

Having a good model of a process is not quite useful if you don't have the appropriate tools to make sense of the resulting representation. For our inferred domains, topics and metadata blocks, we must provide interfaces that properly deploy these abstractions as lenses through which one can interact with and interpret the data, including the relationships between different block types at multiple levels.

The main object of interpretative interest in our approach are documents. They are the material product assembling terms into meaningful texts and connecting the many dimensions of the corpus. Documents get lexically partitioned into domains through a domain-topic model, and those partitions translate to partitions of other dimensions through domain-chained models. It is thus the interpretation of domains, as sets of documents manifesting an assemblage of topics characteristic of the corpus, that will allow us to dissect the data and make sense of its evolution and non-textual dimensions. In this section we present two related interfaces for interpreting domains: domain-topic tables and domain-element maps.

The ability to discuss blocks at different scales, provided by the nested aspect of the model, plays a key role in these interfaces. For that, we will talk of a superdomain and a subdomain to refer to, respectively, the parent and children of a domain in the nested hierarchy. Level-1 domains have no subdomains, as they directly partition documents (see 1D in figure 1). Conversely, the single domain at the highest level has no superdomain, since it contains all documents. The same goes for subtopics and supertopics and, more generally, for superblocks and subblocks of any dimension.

While working at different scales, one must remain attentive to the privileged role of level-1 blocks. For it is in the level-1 domain that we find, manifested in its documents, a concrete and coherent assemblage of topics. Just as it is in level-1 topics that we find terms with a coherent disposition to be employed in some domains but not in others. At higher levels, blocks gather in superblocks, whose connections are the combined connections of their children. But this combined connectivity pattern has no requirement to represent concrete nodes in the data, since, in the generative process of the nested model, concrete links are added exclusively from level-1 probabilities. Therefore, the meaning of higher-level blocks must be built up from their base nodes.

As a consequence, it is thus useful to ask which level-1 topics are distinctive of a level-1 domain, as they can be expected to be articulated together in its documents. But, for higher-level domains, it is more meaningful to ask what they have in common, that is, what topics are distinctive of all of their subdomains. These observations translate to two information theoretical measures that we will employ in our interfaces: the nested specificity and nested commonality.

## Nested specificity and nested commonality

If we consider a domain and its nested superdomains as probability distributions over level-1 topics, corresponding to the usage of topics in the documents they contain, then the domain's distribution can only be positive for topics where the distributions of its parents are positive, a property called absolute continuity. This property allows us to resort to the relative entropy (the Kullback–Leibler divergence) between these distributions in order to quantify the contribution of each topic to the overall information gain incurred if we describe it using the domain's distribution instead of a higher-level superdomain's coarser distribution. In addition, we average the topic's contribution across the superdomain ladder, thus accounting for changes at all scales while still accentuating more local changes. This provides us with an expectation-weighted pointwise nested relative entropy, which will be our nested specificity. Given a domain $d$, its value for a topic $t$ is expressed by the following formula, with $d_+ \supset d$ representing the domains $d_+$ that are superdomains of $d$ at each level above it:

$$\hat{S}_d(t) = \frac{1}{|d_+ \supset d|} \sum_{d_+ \supset d} p_d(t) \log(\frac{p_d(t)}{p_{d_+}(t)})$$

For the nested commonality, we want to find topics that have high specificity in all the subdomains of a domain. We obtain a first quantity by taking the pointwise relative entropy of a topic between each subdomain and a superdomain of the domain, averaging across the domain's superdomain ladder, and then averaging across the subdomains. Given a domain $d$, its value for a topic $t$, with $d_+ \supset d$ as before and $d_- \subset d$ standing for the subdomains $d_-$ of $d$ at the level immediately below it:

$$\hat{C}^*_d(t) = \frac{1}{|d_- \subset d|} \sum_{d_- \subset d} \frac{1}{|d_+ \supset d|} \sum_{d_+ \supset d} \log(\frac{p_{d_-}(t)}{p_{d_+}(t)})$$

This quantity is related to the change, between the compared distributions, in the probability of always sampling that one topic if one samples a topic from each subdomain. It is positive if a topic is overrepresented in all subdomains, and negative if sufficiently underrepresented in at least one subdomain, reaching minus infinity if it is missing from any of them. Note that we take the domain's superdomain ladder in the calculation, to not accentuate the specificity of subdomains towards their own union, which would run contrary to our goal.

We can now account for expectation by considering the term for a topic's contribution to this quantity's expected value, obtaining our nested commonality measure. Consistent with the above, where we account for subdomains as units, we consider a topic's probability as the probability of first uniformly choosing a subdomain, and then sampling a topic from it. We thus have:

$$\hat{C}_d(t) = (\frac{1}{|d_- \subset d|} \sum_{d_- \subset d} p_{d_-}(t)) \ \hat{C}^*_d(t)$$

While we have developed these measures to obtain characteristic topics for domains, the symmetric structure of the nested blocks actually affords repurposing them for any two block types, such as to obtain characteristic domains for metadata blocks, and even characteristic domains for topics.

## Domain-topic table

The domain-topic table answers the need for a simple, static and publication friendly interface. It presents a chosen group of domains, together with the level-1 subdomains forming each of those, and it uses the measures defined above to list the specific topics of the level-1 subdomains, as well as, for each domain in the group, the topics common to their immediate subdomains. For example, if we're interested in a level-3 domain, the domain-topic table lets us examine the group of its level-2 subdomains. For each level-2 domain, it would display what its subdomains have in common, and what is specific to the level-1 domains composing it. In this case, the table can also present what the level-2 domains have in common among themselves, as subdomains of the level-3 domain. Schematically, we have:

| [ level-3 domain ] | | | |
|---|---|---|---|
| [ common topics of its level-2 subdomains ] | | | |
| [ level-2 domain ] | [ common topics ] | [ level-1 domain ] | [ specific topics] |
| | | [ level-1 domain ] | [ specific topics] |
| [ level-2 domain ] | [ common topics ] | [ level-1 domain ] | [ specific topics] |
| | | [ level-1 domain ] | [ specific topics] |

Table 1: Domain topic-table for the level-2 subdomains of a level-3 domain, showing common topics for domains at levels above 1, and specific topics for level-1 domains.

In such a table, domains are represented by a code and label, while topics are represented by a code and a list of terms. In this paper we use the following criteria to choose which topics and terms to display: for topics, we pick the topics with highest specificity or commonality for the domain in question, until the chosen topics account for half of the sum of positive values of that quantity over all topics. For terms, noting that it is straightforward to adapt the specificity and commonality measures by replacing topics with terms, we pick those terms from the topic whose values are higher than half the highest value. The difference in criteria for topics and terms are a consequence of what they represent: distinct topics display different connectivity patterns, so we account for enough topics to cover a majority of the sum of their values; meanwhile, terms within the same topic display similar connectivity patterns, so we simply account for those that contribute the most.

Same as for the measures, these tables can be similarly constructed for other block types, for example, to get a table of characteristic domains for blocks of some metadata dimension.

## Domain-topic map

Domain-topic tables are great when we already have a broader understanding of the corpus and have chosen a set of domains to investigate. But we still lack a tool for interactive exploration of the corpus at different scales and dimensions. This is what domain-topic maps provide.

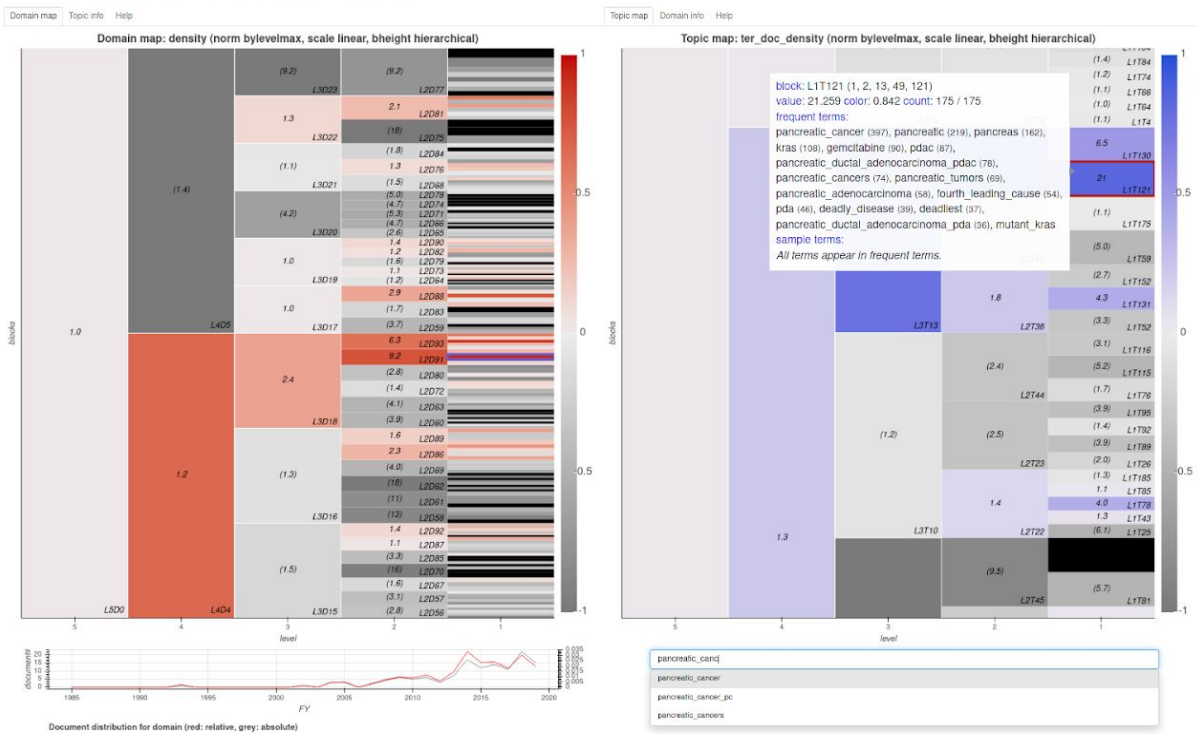

Figure 4: Screen capture of a domain-topic map, with longitudinal histogram and topic search. Domains are shown on the left, in red, and topics on the right, in blue. Columns show the partition at decreasing levels of the nested hierarchy, where each block from a higher level is sliced in sub-blocks of equal height in the lower level. Color intensity tells the relevance of a block for the selected block on the opposite side. In the figure, domains appear colored for their usage of the selected level-1 topic, associated with pancreatic cancers, and topics appear colored for their usage in the selected level-1 domain, which happens to be the domain more strongly associated with the selected topic.

In Figure 4 we see a static frame from a domain-topic map. It displays on its left side the nested structure corresponding to the inferred domains and, on the right side, the nested structure of topics. Color intensity represents the relevance of each block. Upon loading the map, this relevance corresponds, for domains, to the fraction of abstracts found in each domain, and for topics, to the fraction of term usage coming from that topic. In addition, a histogram may be shown under the domain map, graphing the absolute and relative number of documents along some dimension, typically time. The following interactions are then possible:

- cursor over a domain: display the domain's common topics and terms (domain level > 1) or specific topics and terms (level-1 domain), as in the domain-topic table; also display its size and relevance value.
- cursor over histogram: display the document count for the domain and the corpus at that point.
- cursor over a topic: display the most frequently employed terms from the topic; also display its size and relevance value.
- zoom in and out: navigate the hierarchy for closer inspection of low level domains.
- click on a block: change the opposite map to display colors relative to the chosen block, as explained in figure 4; in a new tab, present information and document titles linked to URLs; if a domain, restrict the histogram to its documents.
- search box: equivalent to clicking on a topic, but by searching for a term it contains.

Finally, as before, by simply replacing terms and topics with elements and blocks of some chained metadata dimension, we obtain a domain-chained map, providing the same kind of interactivity between domains and metadata blocks.

In the following sections we'll introduce a dataset and show how navigating these maps lets us identify questions from the emergent block patterns, then seek and consolidate answers by producing specific domain-topic tables and charts.

# 3. Data

## 2.4 The ASCO Annual Meeting abstracts

To test and exemplify our approach, we have analysed a dataset of conference abstracts from the American Society for Clinical Oncology (ASCO) Annual Meeting between 1995 and 2017. Why conference abstracts rather than journal articles? First, because scientific and clinical gatherings such as the ASCO meetings are a major forum for the introduction of the latest clinical and scientific research results, with the understanding that some of those results are preliminary, will not necessarily be confirmed, and will therefore remain unpublished (Massey et al., 2016). Investigating conference abstracts rather than publications thus provides a privileged take on "science in the making" (Latour, 1987), i.e., in our case, the moving front of oncology research, while also opening the possibility of comparing different stages of the production of scientific knowledge.

Second, we're interested in exploring and stressing the fact that, as it requires only the content of individual documents, rather than specific metadata such as citations or coauthorship, our approach is applicable to many kinds of documents, including publications, but also grant proposals, grey literature or, as is our case, conference proceedings. Science studies scholars have rarely investigated scientific conferences. Notable exceptions include Söderqvist and Silverstein (T. Söderqvist & Silverstein, 1994; Thomas Söderqvist & Silverstein, 1994) who analyzed immunology meetings to unravel the sub-disciplinary structure of that discipline: their approach, however, was based on a cluster analysis of meeting participants, rather than an investigation of the content of presentations. A more recent study of scientific conferences has focused on a comparison of ASCO and its European counterpart, ESMO, between 2000 and 2010 (Pentheroudakis et al., 2012). Both conferences are attended, to different degrees, by practitioners from both sides of the Atlantic and other parts of the world. Their analysis concerned presentations of clinical trials, and centered on differences in prevalence, between the two meetings, of characteristics such as industry sponsorship, blinded design, sample size, early interim discontinuation, use of endpoints, and biomarker evaluation.

We also acknowledge an auxiliary issue, equally pertinent to most research on the contents of journal articles: whether abstracts can be considered representative of the content of full presentations or papers. Recent evidence (Ermakova et al., 2018), albeit based on journal abstracts from a different field, shows that abstracts cannot be considered as "mere teasers", and that in fact their "generosity", defined on the basis of the amount and importance of information provided by the abstracts, has been increasing in recent years.

And why ASCO? The ASCO Annual Meeting is the main venue for cancer researchers from around the world to present innovative results. Established in 1964 with 66 members specializing in the then emerging chemotherapy domain, ASCO membership had grown by 2010 to more than 27,000 adherents representing all oncology subspecialties. Its network presently connects close to 45,000 oncology professionals and covers more than 150 countries. Attendance at the annual meeting broke the 5,000 mark in 1985, and in 2018 had increased to 40,700 participants (33,100 professionals and 5,700 exhibitors). Regularly attended by a large number of foreign practitioners (46% of all attendees in 2018), the ASCO Annual Meeting has become one of the largest gatherings of medical professionals in the world. Initially held in conjunction with the annual meeting of the American Association for Cancer Research (AACR), the ASCO conference became self-standing in 1993, and, according to its organizers, by 1995 a decision was made to increase its emphasis on translational science. We can thus safely claim that papers selected for presentation at ASCO provide a representative sample of oncology investigations at the international research front, often including practice-changing results.

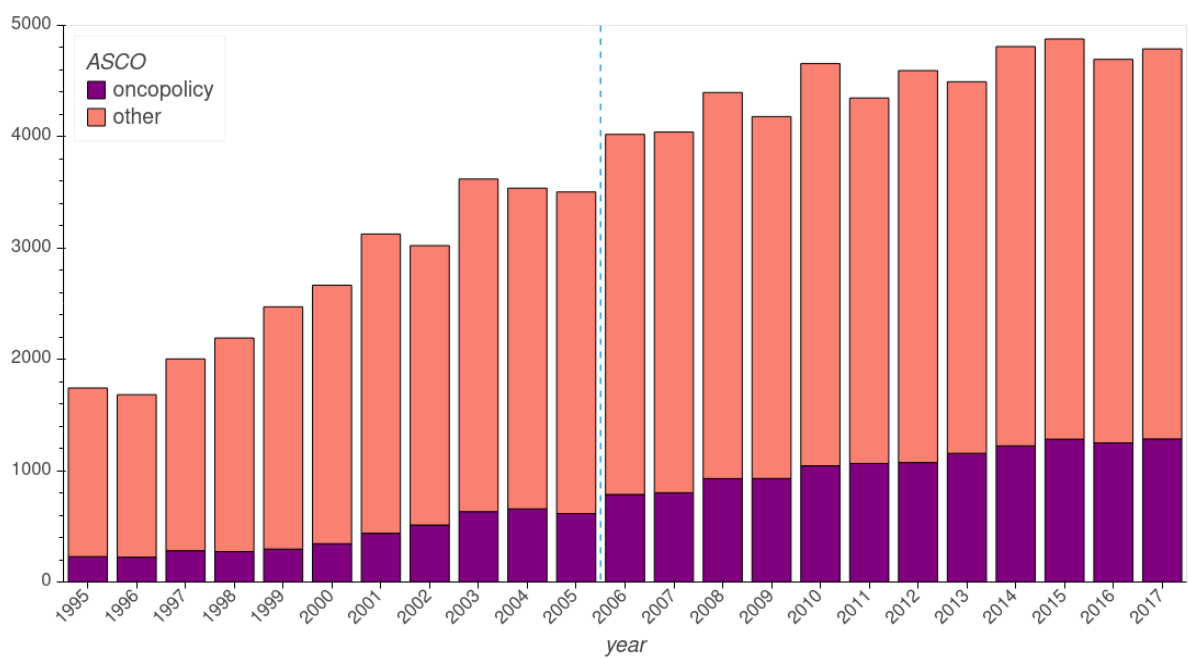

Figure 5: Yearly volume of the corpus of abstracts from the American Society for Clinical Oncology (ASCO) Annual Meeting between 1995 and 2017, totaling 83,476 abstracts. Advancing some of the results of our analysis, we highlight the contribution of a group of domains we'll label "oncopolicy", and show the split of the two main periods to be inferred from the data.

Figure 5 shows the evolution of the number of abstracts presented at the ASCO Annual Meeting from 1995 to 2017. After a period of steady growth, the yearly number of abstracts reached a plateau of approximately 5000 abstracts in the 2010s. There are both practical and theoretical reasons for our decision to examine these abstracts and select this 23-year window. Oncology has played a pioneering role in the much-touted area of translational research (Cambrosio et al., 2006), a subfield characterized by rapid change, and the ASCO Annual Meeting is the main annual event in this domain. As just mentioned, 1995 coincides with a readjustment of the socio-technical content of the annual meetings that continues to characterize them to the present day. We can therefore test our approach on a reasonably stable, coherent set of documents that pertain, at the same time, to a dynamic domain. The highly selective nature of the ASCO Annual Meeting will moreover dispense us with the thorny issue of separating the grain of what practitioners consider innovative contributions from the chaff of the mundane or routine findings that can be found in a rapidly expanding number of oncology journals.

Finally, we should note that presentations at ASCO meetings are distributed throughout different conference sections and sessions according to their subject. However, these "native" categories are primarily designed for organizational purposes, and thus highly contingent. By contrast, we focus on the processing of the abstracts' raw textual material as it provides a direct take on the thematic landscape of the conference as a whole and across time.

## 2.5 Data processing

We now turn to a more technical description of the processing pipeline we apply to the ASCO abstracts in preparation to infer their domain-topic block structure. To emphasize the possibilities of the model, we adopt a procedure that is entirely language independent. We intentionally do not filter stop-words, and we do not apply any language-specific NLP procedure such as stemming or lemmatizing. But we still search for co-location using classical statistical analysis (Mikolov et al., 2013) to extract frequent bigrams in the corpus such as *"stem_cells"*, *"breast_cancer"*, *"partial_response"*, *"aplastic_anemia"*, or "*colorectal_cancer*". We then index each document by its own set of terms, i.e. words and bigrams, effectively transforming each abstract into a vector that codes for the presence or absence of each term, thus following the classical bag-of-words prescription (Turney & Pantel, 2010) to model documents.

We also differ in two important ways from the more traditional text-analysis procedure (Blei, 2012): (i) we do not *a priori* filter out very rare (singletons) or frequent (stop-words) terms, to emphasize how the model is capable of effectively treating any connectivity pattern; and (ii) our document vectors are binary, meaning that we only account for the presence or absence of a term in a document. Regarding (ii), by ignoring the local frequency of terms we put more weight on the thematic features of the text and avoid patterns based on stylistic variations like preferences to replace nouns by pronouns. We do not argue that these are the best choices, only that they make sense in our case of quite uniform documents, and for our purpose of illustrating what can be achieved with no dependency of additional procedures.

# 4. Results

To facilitate our discussion of multiple domains and topics, we adopt the notation below to refer to an individual domain or topic block at a particular level:

$$L_iD_j \equiv \text{domain } j \text{ at level } i$$

$$L_iT_j \equiv \text{topic } j \text{ at level } i$$

So, for instance, L2T29 is the topic with index 29 at level 2, and L3D40 is the domain with index 40 at level 3. Although this notation distinguishes domains from topics, all blocks are indexed together in such a way that, for a given level, domain indices start after the highest topic index.

To begin our analysis of the ASCO Annual Meeting abstracts, we produced a domain-topic model of its contents and then a domain-chained model of the meeting years associated with each abstract. For the domain-topic model, Table 2 shows its block counts, and we also produce a domain-topic map to work with (in SI-MAPS). For the domain-chained model, because the number of years is small, we directly represent the nested blocks pictorially in Figure 6, which we call call periods. We also produce a domain-period map to work with (in SI-MAPS).

|         | L5 | L4 | L3 | L2  | L1  | N      |
|---------|----|----|----|-----|-----|--------|
| Domains | 1  | 4  | 24 | 110 | 479 | 83476  |
| Topics  | 1  | 5  | 37 | 112 | 407 | 253758 |

Table 2: Partition counts for domains and topics at each nested level of the domain-topic model, plus the number (N) of documents and terms.

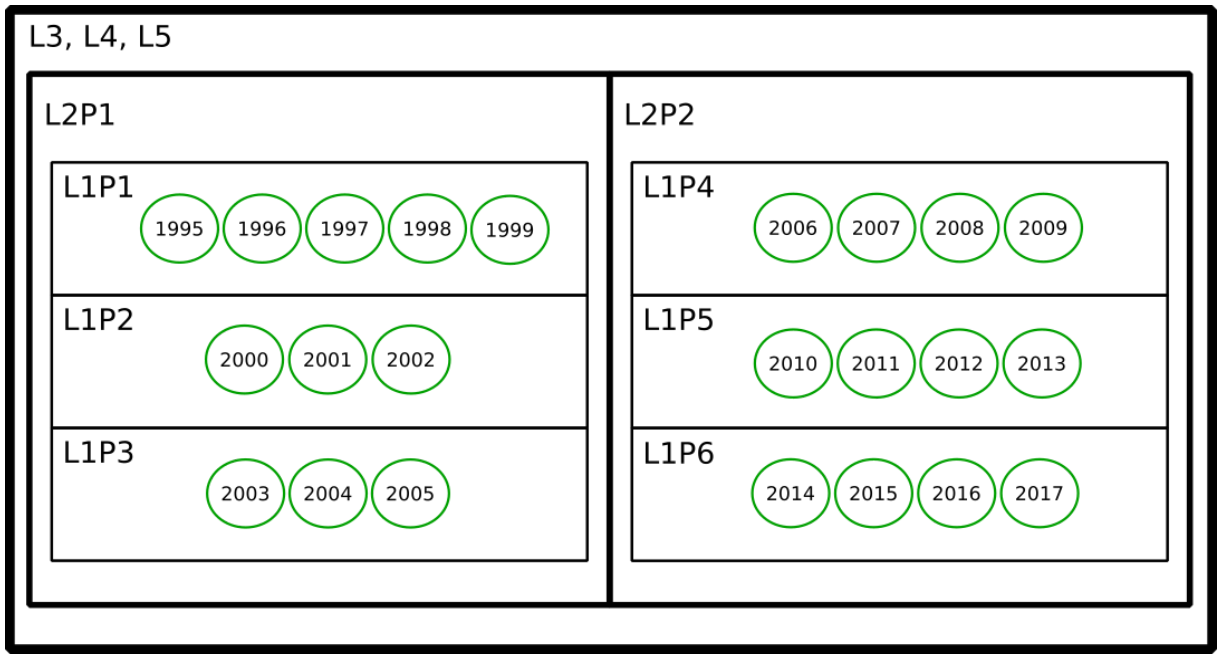

Figure 6: The nested partition of annual meetings into periods. Contrary to domains and topics, levels L3, L4 and L5 of conference years are all equivalent, as the inference procedure found no statistically significant distinctions above level L2. Moreover, since conferences are treated as categorical data, the fact that the partitions respect the chronological sequence is not a given, but reveals a progressive character in the evolution of research domains at the ASCO Annual Meetings.

We then proceeded by navigating the domain-period map, in order to understand what are the research domains going through meaningful shifts in their prevalence between the periods identified. We can also formalize this notion, by taking the difference of a domain's prevalence between subsequent periods. In this case, by considering the two main periods, [1995-2005]$^{L2P1}$ and [2006-2017]$^{L2P2}$, we obtain the color map in Figure 7 showing the growth or decline of domains between them, to which we've added labels for some notable domains. At level-1, the block with the highest growth refers to survival and prognosis across different cancers$^{L1D808}$, while the block with the strongest decline refers to traditional chemotherapy for lung cancer$^{L1D458}$ and is consistent with the demise of cytotoxic chemotherapy approaches at the research front. At level-2, the highest growth corresponds to cancer genomics as defined by work on genomic alterations, mutations, and molecular profiles across cancer types$^{L2D133}$.

We also find a very stable domain that corresponds to hereditary cancer[L2D139]. And finally, at level-3, a high-growth domain that brings together fields like public health, healthcare policy, healthcare services, and cost analysis[L3D44], i.e. a number of miscellaneous contributions to which oncologists increasingly refer using the umbrella term "oncopolicy"

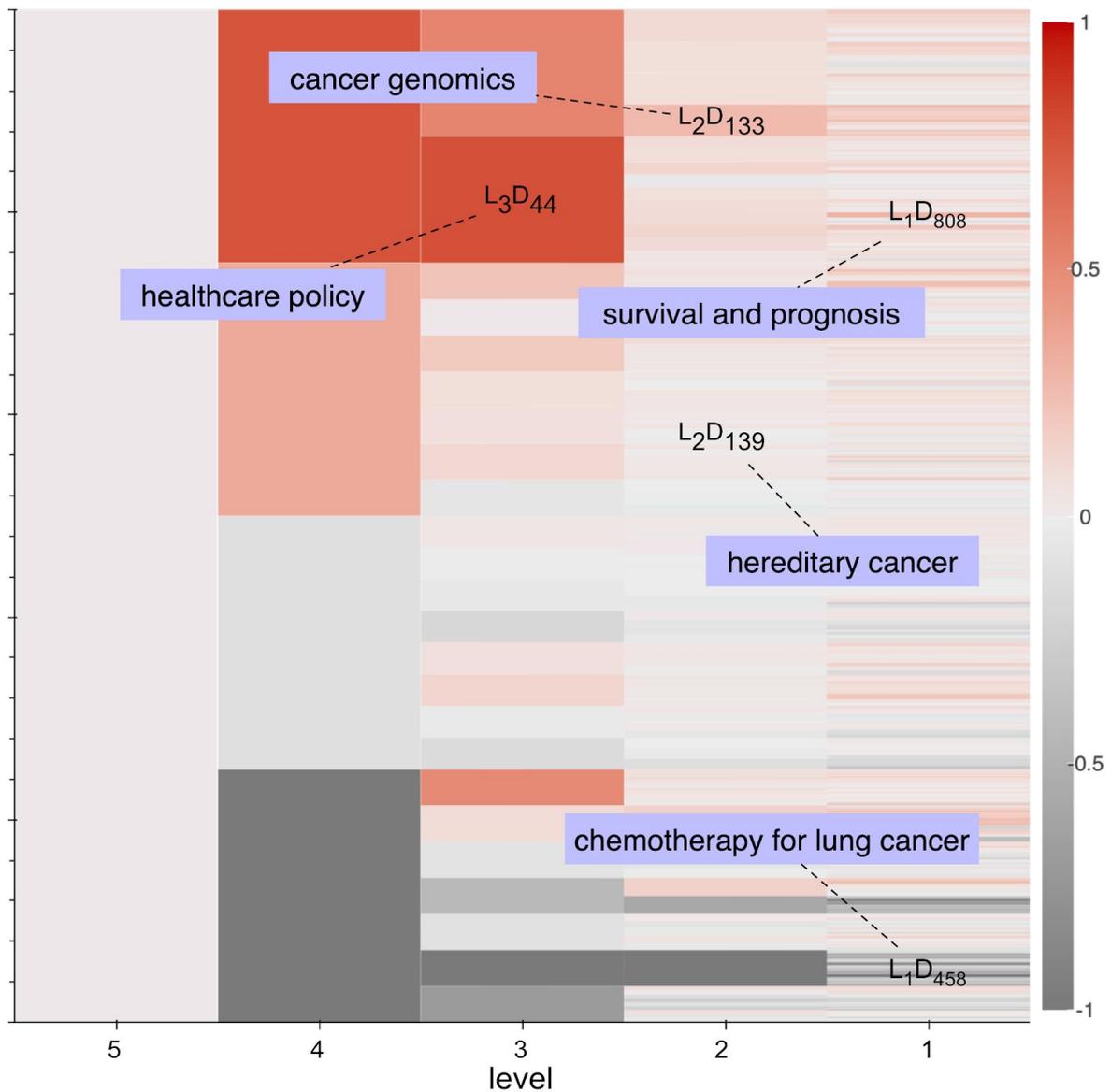

Figure 7: Colors represent the growth, in red, or decline, in grey, of the prevalence of a domain, that is, the fraction of abstracts contained in it for each period. We consider the evolution between the two highest-level periods: [1995-2005] and [2006-2017]. Colors are normalized within each level, so that the blocks of strongest color correspond to the greatest change at their level of granularity. Labels are the result of procedures akin to what we'll perform for L3D44.

We note that these maps offer the opportunity to delve into the details of the whole corpus. They afford both, to perform an exhaustive analysis, and to seek guidance in a concrete question and proceed to uncover the relevant domains. Here we've been guiding ourselves by the shifts in prevalence, after having paritionned the temporal dimension. Among those domains of notable shift in relevance, we find one deserving particular attention, as it reveals the growing importance of social, economic and political issues that weren't always central to the meetings. And it actually corresponds to a major reorientation currently taking place not only at at ASCO, but also within other oncology organizations such as the National Comprehensive Cancer Network (McNeil, 2018), according to which health policy topics that a few years ago would not have been discussed at meetings have become legitimate topics given the present "era of rapid change — in therapies, costs, payment models, and practice".

In fact, in early 2020 ASCO launched a "professional association, that will enable expanded advocacy activities and increase the impact of efforts directed toward policymakers in support of high-quality patient care." (ASCO, 2020). Similar policy initiatives are to be found within European oncology organizations. For the remaining of this paper we will thus unravel the domain coded as L3D44.

L3D44 has 10 level-2 subdomains, each one having its own set of level-1 domains. Our starting goal is to label the 10 subdomains by studying the full domain-topic table for L3D44. If we're successful in finding, for each level-2 domain, a label that accounts for the abstract coherence of its set of level-1 subdomains, we'll not only have understood our domain in terms of 10 meaningful subdomains, but we can then ask how those ten domains are related to other dimensions. To produce these labels, we employ the domain-topic table for domain

L3D44 (in SI-TABLES). It is, however, impractical to include the full table here, as it contains about a hundred level-1 domains. To present the procedure, we will show an extract of the table for one level-2 domain, and then a simplified table containing the 10 labels but omitting the level-1 details required to formulate them.

Table 3 presents the domain-topic table for L2D120. As explained in Section 2.3, individual level-1 domains can be more directly interpreted as they are concretely coherent, with their assemblage of topics manifested in the documents they contain. By inspecting and interpreting this ensemble of level-1 domains, we arrive at an appropriate label for their parent domain: "quality of life and treatment side-effects".

| L2 | Common Topics | L1 | Specific Topics |
|---|---|---|---|
| D120 | • T156 (9%): **repeated_measures** (100%)<br><br>• T67 (8%): **questionnaire** (76%)<br><br>• T14 (8%): **scores** (28%), score (20%), scale (16%)<br><br>• T272 (7%): **baseline** (45%)<br><br>• T23 (6%): **pain** (35%), symptoms (34%), severity (20%)<br><br>• T241 (4%): **randomized** (50%)<br><br>• T12 (4%)<br><br>• T249 (3%): **quality** (54%), life (40%)<br><br>• T367 (3%): **measured** (38%), effect (33%), effects (24%) | D415 | T156 (23%): **qol** (3%), scales (3%), life_qol (2%), subscales (2%), internal_consistency (2%) • T67 (15%): **items** (6%), item (3%), questionnaire (3%) • T14 (10%): **scale** (12%), scores (9%), validity (9%), reliability (8%), measure (8%) • T23 (4%): **symptoms** (17%), symptom (14%), pain (13%) |
| | | D563 | T67 (42%): **exercise** (2%), depression (2%), physical_activity (2%), physical (1%), anxiety (1%) • T249 (5%): **intervention** (32%) • T156 (4%): **qol** (13%), life_qol (8%) |
| | | D584 | T156 (30%): **qol** (6%), eortc_qlq-c30 (4%) • T14 (6%): **scores** (24%), score (16%) • T67 (5%): **questionnaire** (6%), physical (6%), questionnaires (5%), items (3%) • T241 (4%): **arms** (19%), arm (17%), randomized (17%) • T272 (4%): **baseline** (41%) • T23 (3%): **symptoms** (17%), pain (11%), deterioration (10%) |
| | | D614 | T134 (54%): **bone** (4%), bone_metastases (4%), zoledronic_acid (3%), bone_resorption (2%) |
| | | D635 | T38 (54%): **weight** (3%), bmi (3%), body_composition (2%), cachexia (2%), nutritional_status (2%) |
| | | D687 | T12 (28%): **cipn** (3%), chemotherapy-induced_peripheral (2%), neuropathy_cipn (2%), hands (2%), feet (1%) • T23 (4%): severity (18%), pain (16%) • T79 (4%): **neuropathy** (13%), peripheral_neuropathy (9%), neurotoxicity (7%) • T142 (3%): placebo (30%) • T181 (3%): **skin** (14%), topical (10%), cutaneous (9%) • T241 (3%): **randomized** (21%), arm (16%), arms (15%) • T28 (2%): **daily** (22%) • T283 (2%): **side_effects** (20%), severe (17%), side_effect (17%) • T355 (2%): incidence (24%), prevention (24%) |
| | | D805 | T40 (19%): **approval** (4%), fda (4%) • T228 (5%): **drug** (38%) • T93 (3%): **approved** (15%) • T65 (3%): ae (9%), aes (8%), sae (5%), saes (5%) • T251 (3%): **design** (26%) • T236 (3%): **review** (14%), reports (13%), reporting (10%), published (7%) • T292 (2%): **trials** (59%) • T29 (2%): **maximum_tolerated** (14%), mtd (12%), dlt (9%), dose_escalation (9%) • T11 (2%): **supplemental_indications** (2%), marketing (2%), notification (1%), biopharmaceutics (1%), budgeted (1%) • T395 (2%): **phase** (81%) • T258 (2%): **clinical_trials** (72%) • T240 (2%): **safety** (29%) • T239 (2%): new (35%) • T230 (1%): accrual (32%) |

Table 3: Line L2D120 from the domain-topic table for L3D44. Percentages are the item's fraction of total positive commonality or similarity contributions, either at the topic or term level, as defined in the text. Topics are always level-1. Note that L1T12 has no terms present in all subdomains of L2D120.

Going through this process for each of the 10 subdomains of L3D44 yields the labels presented in Table 4, which also provides some summary information on each level-2 domain. Then, by looking at the ensemble of these level-2 labels, we can provide a label for the whole of the level-3 domain: "health technology assessment (broadly understood to include treatment regimens)", which we identify with the native yet vague term "oncopolicy".

| | **L3D44** | | |
|---|---|---|---|
| | **Label:** Health technology assessment (broadly understood to include treatment regimens) | | |
| | **Common Topics:** T127: **health_care**, physicians, physician, community · T343: **care**, medical, health · T58: **utilization**, claims · T135: **more_likely**, adjusted, demographic, logistic_regression · T33: **less_likely**, receipt · T260: **use** · T72: **hospital**, national, center · T102: **should_be**, regarding, they, should · T222: **practice**, guidelines · T325: **over**, period · T96: **issues**, how, clinic, needs, participants · T15: **age**, older, mean_age · T249: **quality**, life · T84: **cost**, costs · T304: **outcomes**, impact · T150: **comorbidities**, comorbidity · T198: **among**, those · T116: **hospitalization**, outpatient, palliative_care, inpatient, length · T187: **more**, most · T14: **score**, measures, scores · T273: **population**, us, proportion · T341: **diagnosis**, diagnosed · T215: **clinicians**, actual, preferred, decision, preference · T170: **compared**, than, lower, greater, higher | | |
| | **L2 Subdomains** | | |
| **L2** | **Label** | **N** | **Summary** |
| D120 | *Quality of life and treatment side effects* | 1796 | Quality of life (physical and psychological). Treatment side effects. |
| D121 | *Quality improvement* | 2303 | Quality improvement (professionals, costs, and practices). Professional education and communication with patients. |
| D124 | *Evaluation of treatment regimens* | 1343 | Comparative evaluation of treatment regimens (resources, costs, efficacy, side effects). |
| D126 | *Epidemiological surveillance of outcomes* | 1931 | Epidemiological surveillance of outcomes (SEER = Surveillance, Epidemiology, and End Results). Prognosis. |
| D131 | *Risks and management of treatment side effects* | 1387 | Risks and management of treatment side effects (toxicity, infection, etc.) |
| D147 | *Assessment of physical and psychosocial side effects* | 1523 | Assessment of physical and psychosocial side effects. Hospitalization. |
| D151 | *Pain management and quality of life* | 805 | Pain management and quality of life (randomized studies thereof). |
| D153 | *Survey questionnaires of quality of life and professional adherence to guidelines* | 1844 | Survey questionnaires of quality of life (including depression and anxiety). Survey of professional adherence to clinical guidelines. |
| D163 | *Patient lifestyle education* | 1142 | Patient education concerning lifestyles. Barriers to screening. |
| D186 | *Meta-analysis of clinical trials* | 953 | Meta-analysis of clinical trials (all sorts of cancers). |

Table 4: Complete description of the "health technology assessment"[L3D44] domain in terms of its common topics and subdomains. Level-2 subdomains were labeled by employing the domain-topic table for the level-3 domain, whose row describing the "quality of life and treatment side effects"[L2D120] subdomain was presented as table 3.

Still, one may ask if we're not losing some meaningful related material by focusing only on this single level-3 domain. To answer that, we look at the topics common to its subdomains, and notice the main shared topic is L1T127, whose most shared term is "health_care". By going back to the domain-topic map and selecting that topic, it becomes clear that this domain dominates the usage of L3D44. Yet a single other level-3 domain still stands out: L3D48. By similarly inspecting the next topics for the first domain, we can see that this other domain does also fit within the realm of oncopolicy, and proceed to include it in our analysis. The resulting labels and summary information are presented in Table 5.

| L3D48 |
|---|
| **Label:**<br>screening and risk factors for cancers (in particular hereditary and secondary cancers), including epidemiological surveillance |
| **Common Topics:**<br>T80: **screening** · T97: **family_history** · T73: **seer**, surveillance_epidemiology, white, seer_database, race · T33: **population-based**, registries · T341: **diagnosis**, diagnosed, registry · T109: **mortality**, general_population · T3: **country**, countries, world · T72: **national**, center · T96: **participants** · T127: **health_care**, community, physicians, referral, physician · T78: **survey**, education · T15: **age** · T246: **risk** · T47: **invasive**, mammography · T273: **population** · T222: **guidelines**, recommendations · T242: **cancer**, cancers · T362: **breast_cancer**, breast |
| **L2 Subdomains** |

| L2 | Label | N | Summary |
|---|---|---|---|
| D134 | *Breast cancer screening and prophylaxis* | 588 | Screening (mammography and genetic testing) for breast cancer, esp. hereditary one, and prophylaxis, including practice guidelines. |
| D139 | *Genetic testing and counselling for hereditary cancer risks* | 509 | Genetic testing and genetic counselling for hereditary cancer risks. Epidemiological surveillance. |
| D141 | *Cancer risk factors* | 433 | Cancer risk factors, including race, ethnicity, smoking, and obesity. |
| D159 | *Epidemiology of cancer, esp. blood cancers* | 426 | Epidemiology of cancer, esp. blood cancers. Risk factors for secondary cancer (following treatment of primary cancer). Epidemiological surveillance in different world populations. |
| D180 | *Molecular pathology: quality control, technology comparisons, and regulation* | 342 | Molecular pathology: quality and comparison of different technologies. Molecular testing and its regulation, access to testing, expertise and decision support for testing (both human expertise and computational approaches). |

Table 5: Complete description of the "screening and risk factors for cancers"$^{\text{L3D48}}$ domain.

Having defined our perimeter of interest as the union of domains "health technology assessment"$^{\text{L3D44}}$ and "screening and risk factors for cancers"$^{\text{L3D48}}$ – which henceforth we'll call the "oncopolicy domains" or simply "oncopolicy" – and having carefully labeled their subdomains in such a way that they provide us with a set of meaningful and related, yet distinctive aspects of our subject of interest, we are now enabled to depict and describe its evolution at a more detailed level.

Figure 8 shows an area bump chart for all 15 level-2 subdomains within "oncopolicy", across the 6 level-1 periods previously inferred (see figure 6). While they're growing in absolute terms, relative to each other we witness a loss of relevance for the pain management[L2D151] and breast cancer screening[L2D134] subdomains, accompanied by an expansion of those subdomains concerned with treatment regimens[L2D124], metanalysis[L2D186] and molecular pathology[L2D180]. At the same time, the larger subdomains have remained rather stable throughout these periods, despite "quality of life and treatment side effects"[L2D120] lagging a bit behind and "assessment of physical and psychosocial side effects"[L2D147] showing up, even as "oncopolicy" developed into a major topic within ASCO.

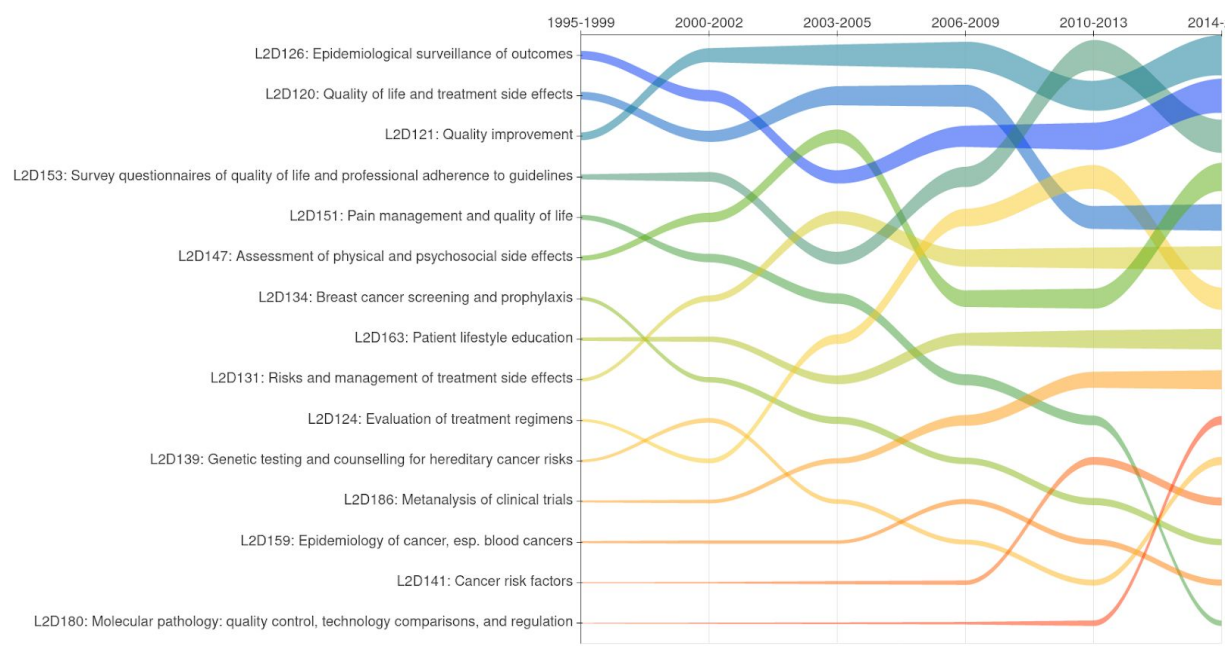

Figure 8: Area bump chart for "oncopolicy" (L3D44 and L3D48) in ASCO, with its 15 level-2 domains displayed on the y-axis, highlighting rank and volume changes along the 6 level-1 periods. While most domains grow in time, relative to each other some domains have seen a steep decline (notably L2D151 and L2D134), while others have noticeably mounted in rank and scale (e.g. L2D186 and L2D180). Volumes are year averages within each period.

Before we conclude our analysis, we very briefly describe, without thoroughly presenting these results, a few more procedures afforded by our approach: first, we can take the "oncopolicy" domain and treat it as a corpus on its own, replaying the domain-chained model of time to obtain a new partition into periods that reflect only the contents of "oncopolicy", and not the whole of ASCO. This yields a simpler partition into only three blocks (1995-2005, 2006-2012, 2013-2017). We can then, for example, take the later of these oncopolicy-specific periods to represent its more current research trends, and restrict once more the corpus to only include such recent documents ("oncopolicy" and [2013-2017]). And, on top of this subcorpus, we can run a domain-chained model of the countries found in the affiliation metadata, to cluster them from their insertion in "current oncopolicy". Of which, by employing a country-domain table, among other features we can observe how the bulk of European countries, together with Brazil, form a single cluster displaying a slight tendency towards "Metanalysis of clinical trials"$^{L2D186}$ and "Quality of life and treatment side effects"$^{L2D120}$.

## 5. Discussion

We have shown how domain-topic models and domain-chained models, allied with measures, tables and maps tailored to render the models' state representations accessible to human researchers, afford a fine-grained and multifaceted view over the scientific dynamics of the ASCO Annual Meetings. The domain-topic model formed the basis of our approach, abstracting topic-wise similar documents into domains, and domain-wise similar terms into topics, at different levels of scale. It systematized and facilitated the navigation and characterisation of the corpus.

Then, applying a domain-chained model to the meeting's years allowed us to identify content based periods, whose consistent chronology revealed the conference's progressive dynamics. Through these periods, we located and broadly described a few shifting domains. Among which, a growing level-3 domain of interest to major current events in ASCO and other oncology organizations, related to "health policy". We chose to focus on this issue, and by inspecting the topics of this domain we located and expanded our scope to another, strongly related level-3 domain. We then analysed this group of two domains through their multiple scales of subdomains, allowing us to characterise first each subdomain, then the domains, and finally the ensemble, which we labeled "oncopolicy". Following that, by crossing the periods with these characterised subdomains, we provided a granular description of the evolution and trends of "oncopolicy". And, lastly, we have briefly introduced how one may employ sequences of operations with domain-chained models over metadata dimensions, and the production of targeted sub-corpora, in order to perform more complex inquiries into the data.

In line with a recent special issue of Quantitative Science Studies (Leydesdorff et al., 2020) that called for a new alliance between computational and qualitative investigations of techno-scientific activities, our goal has been to show one way in which advanced statistical techniques, deployed in a modular fashion, may provide rigorous yet flexible abstractions that enable original insights, which can be complemented by qualitative investigation and fed back into the inquiry, in a disciplined cycle. Moreover, and in the spirit of providing a vocabulary contributing to the establishment of such trading zones between computational and qualitative approaches, one can argue that domains, as defined in this paper, could be equated, in more qualitative terms, to epistemic communities (Akrich, 2010), whereby topics would amount to the discursive resources mobilized by those epistemic communities, thus acting as bridges between them.

This work leaves a few questions to be addressed in further research, both regarding the approach and the corpus studied. For the approach, although we make a strong case for our choice of model, one may ask whether other models, such as those discussed in the introduction, could be expanded to perform its role, and so a quantitative or systematic approach for comparison with alternatives is desirable. Another interesting avenue would be to consider small corpora that do not generate a detailed topic partition on their own, but may borrow a reference topic partition from a larger and thematically related corpus, which could afford the discovery of domains in the small corpus by employing the same procedure as the chained-model presented here. For the corpus, it would be interesting to compare the ASCO Annual Meeting with other conferences such as ESMO or the smaller ASCO meetings, by modeling them together and observing, for example, their differential insertion in domains. Beyond that, studying conferences, such as the ASCO Annual Meeting, together with related grant proposals and publications, may provide an original view into research cycles as long as one can account for the diversity of these corpora and their metadata, a task for which our approach seems an able candidate. We conclude with a note, that the methods introduced, while built around the study of scientific activities, may be applicable to the study of other subjects, so long as the subject is associated with a corpus whose documents play a connective role across its dimensions.

# Code and data availability

A software library that allows one to perform the steps described in this paper is available at <https://gitlab.com/solstag/abstractology/> as FLOSS (GPLv3). Our source code makes use of the graph-tool library (Peixoto, 2014a).

The American Society of Clinical Oncology (ASCO) retains the rights to the raw data. Those interested in obtaining the full dataset for research purposes should contact ASCO. See <https://www.asco.org/research-progress/asco-data-library> for more details. The authors have made post-processed data available as supporting information (in SI-DATA), containing: for each document, the domain and period to which it belongs; for each term, the topic to which it belongs; and for each topic, a list of the documents in which it can be found. These data allow readers to use the inferred structure and conduct further analysis at the same level of detail discussed in this paper. Readers must obtain the full dataset from ASCO to infer the domain-topic structure, as it requires the content of abstracts.

# Author contributions

Alexandre Hannud Abdo: Conceptualization, Data curation, Formal analysis, Investigation, Methodology, Resources, Software, Validation, Visualization, Writing – original draft, Writing – review & editing. Jean-Philippe Cointet: Conceptualization, Funding acquisition, Methodology, Resources, Validation, Writing – original draft, Writing – review & editing. Pascale Bourret: Conceptualization, Methodology, Resources, Validation, Writing – original draft, Writing – review & editing. Alberto Cambrosio: Conceptualization, Funding acquisition, Methodology, Project administration, Resources, Supervision, Validation, Writing – original draft, Writing – review & editing.

# Acknowledgments

We thank our Co-PI James A. Evans (University of Chicago) and his assistant Antonio Nanni for their suggestions and assistance, as well as ASCO, and in particular Dr. Richard L. Schilsky, for providing access to electronic copies of the annual conference abstracts. We also thank Gil Eyal (Columbia University), Moran Levi (U. Bielefeld) and their assistant Juan L. Martin for helpful discussions.

# Funding information

Work for this paper has been made possible by a grant from the Canadian Institutes of Health Research (CIHR - MOP-142478).

# Ethical review

The work presented here has received ethical approval by the Institutional Review Board of the Faculty of Medicine of McGill University (IRB Study Number A07-E55-15B).

# Competing interests

The authors declare there to be no competing interests. The funder had no part in designing the study or describing the results.